\documentclass[superscriptaddress,aps,prd,twocolumn,altaffilletter,bibnotes,floatfix]{revtex4}
\usepackage[dvips]{graphicx,color} 
\usepackage{array,hhline,dcolumn} 
\usepackage{rotating} 

\newcommand{\kov}{$\check\mathrm{C}$erenkov }

\newcommand{\numu}{\nu_{\mu}}

\newcommand{\nue}{\nu_e}

\newcommand{\pizero}{\pi^{0}}

\begin{document}


\preprint{MiniBooNE-muonbrem}

\title{Constraining Muon Internal Bremsstrahlung as a Contribution to the MiniBooNE Low Energy Excess}

\newcommand{\bama}{University of Alabama; Tuscaloosa, AL 35487}
\newcommand{\bucknell}{Bucknell University; Lewisburg, PA 17837}
\newcommand{\cinci}{University of Cincinnati; Cincinnati, OH 45221}
\newcommand{\colorado}{University of Colorado; Boulder, CO 80309}
\newcommand{\columbia}{Columbia University; New York, NY 10027}
\newcommand{\embry}{Embry-Riddle Aeronautical University; Prescott, AZ 86301}
\newcommand{\fnal}{Fermi National Accelerator Laboratory; Batavia, IL 60510}
\newcommand{\florida}{University of Florida; Gainesville, FL 32611}
\newcommand{\indiana}{Indiana University; Bloomington, IN 47405}
\newcommand{\lanl}{Los Alamos National Laboratory; Los Alamos, NM 87545}
\newcommand{\lsu}{Louisiana State University; Baton Rouge, LA 70803}
\newcommand{\umich}{University of Michigan; Ann Arbor, MI 48109}
\newcommand{\princeton}{Princeton University; Princeton, NJ 08544}
\newcommand{\marys}{Saint Mary's University of Minnesota; Winona, MN 55987}
\newcommand{\vtech}{Virginia Polytechnic Institute \& State University; Blacksburg, VA 24061}
\newcommand{\yale}{Yale University; New Haven, CT 06520}

\newcommand{\beijing}{Institute of High Energy Physics; Beijing 100049, China}
\newcommand{\hope}{Hope College; Holland, MI 49423}
\newcommand{\iit}{Illinois Institute of Technology; Chicago, IL 60616}
\newcommand{\imsa}{Illinois Mathematics and Science Academy; Aurora IL 60506}
\newcommand{\massit}{Massachusetts Institute of Technology; Cambridge, MA 02139}
\newcommand{\caltech}{California Institute of Technology; Pasadena, CA 91125}
\newcommand{\bu}{Boston University; Boston, MA 02215}
\newcommand{\valencia}{IFIC, Universidad de Valencia and CSIC; 46071 Valencia, Spain}
\newcommand{\triumf}{TRIUMF; Vancouver, BC, V6T 2A3, Canada}
\newcommand{\imperial}{Imperial College; London SW7 2AZ, United Kingdom}

\affiliation{\bama}
\affiliation{\bucknell}
\affiliation{\cinci}
\affiliation{\colorado}
\affiliation{\columbia}
\affiliation{\embry}
\affiliation{\fnal}
\affiliation{\florida}
\affiliation{\indiana}
\affiliation{\lanl}
\affiliation{\lsu}
\affiliation{\umich}
\affiliation{\princeton}
\affiliation{\marys}
\affiliation{\vtech}
\affiliation{\yale}

\author{A.~A. Aguilar-Arevalo}\affiliation{\columbia}
\author{A.~O.~Bazarko}\affiliation{\princeton}
\author{S.~J.~Brice}\affiliation{\fnal}
\author{B.~C.~Brown}\affiliation{\fnal}
\author{L.~Bugel}\affiliation{\columbia}
\author{J.~Cao}\altaffiliation{Present Address: \beijing}\affiliation{\umich}
\author{L.~Coney}\altaffiliation{Present Address: \hope}\affiliation{\columbia}
\author{J.~M.~Conrad}\affiliation{\columbia}
\author{D.~C.~Cox}\affiliation{\indiana}
\author{A.~Curioni}\affiliation{\yale}
\author{Z.~Djurcic}\affiliation{\columbia}
\author{D.~A.~Finley}\affiliation{\fnal}
\author{B.~T.~Fleming}\affiliation{\fnal}\affiliation{\yale}
\author{R.~Ford}\affiliation{\fnal}
\author{F.~G.~Garcia}\affiliation{\fnal}
\author{G.~T.~Garvey}\affiliation{\lanl}
\author{C.~Green}\affiliation{\lanl}\affiliation{\fnal}
\author{J.~A.~Green}\affiliation{\indiana}\affiliation{\lanl}
\author{T.~L.~Hart}\altaffiliation{Present Address: \iit}\affiliation{\colorado}
\author{E.~Hawker}\altaffiliation{Present Address: \imsa}\affiliation{\lanl}\affiliation{\cinci}
\author{R.~Imlay}\affiliation{\lsu}
\author{R.~A. ~Johnson}\affiliation{\cinci}
\author{P.~Kasper}\affiliation{\fnal}
\author{T.~Katori}\affiliation{\indiana}
\author{T.~Kobilarcik}\affiliation{\fnal}
\author{I.~Kourbanis}\affiliation{\fnal}
\author{S.~Koutsoliotas}\affiliation{\bucknell}
\author{E.~M.~Laird}\affiliation{\princeton}
\author{J.~M.~Link}\affiliation{\columbia}\affiliation{\vtech}
\author{Y.~Liu}\affiliation{\umich}
\author{Y.~Liu}\affiliation{\bama}
\author{W.~C.~Louis}\affiliation{\lanl}
\author{K.~B.~M.~Mahn}\affiliation{\columbia}
\author{W.~Marsh}\affiliation{\fnal}
\author{P.~S.~Martin}\affiliation{\fnal}
\author{G.~McGregor}\affiliation{\lanl}
\author{W.~Metcalf}\affiliation{\lsu}
\author{P.~D.~Meyers}\affiliation{\princeton}
\author{F.~Mills}\affiliation{\fnal}
\author{G.~B.~Mills}\affiliation{\lanl}
\author{J.~Monroe}\altaffiliation{Present Address: \massit}\affiliation{\columbia}
\author{C.~D.~Moore}\affiliation{\fnal}
\author{R.~H.~Nelson}\affiliation{\colorado}
\author{P.~Nienaber}\affiliation{\marys}
\author{S.~Ouedraogo}\affiliation{\lsu}
\author{R.~B.~Patterson}\altaffiliation{Present Address: \caltech}\affiliation{\princeton}
\author{D.~Perevalov}\affiliation{\bama}
\author{C.~C.~Polly}\affiliation{\indiana}
\author{E.~Prebys}\affiliation{\fnal}
\author{J.~L.~Raaf}\altaffiliation{Present Address: \bu}\affiliation{\cinci}
\author{H.~Ray}\affiliation{\lanl}\affiliation{\florida}
\author{B.~P.~Roe}\affiliation{\umich}
\author{A.~D.~Russell}\affiliation{\fnal}
\author{V.~Sandberg}\affiliation{\lanl}
\author{R.~Schirato}\affiliation{\lanl}
\author{D.~Schmitz}\affiliation{\columbia}
\author{M.~H.~Shaevitz}\affiliation{\columbia}
\author{F.~C.~Shoemaker}\affiliation{\princeton}
\author{D.~Smith}\affiliation{\embry}
\author{M.~Sorel}\altaffiliation{Present Address: \valencia}\affiliation{\columbia}
\author{P.~Spentzouris}\affiliation{\fnal}
\author{I.~Stancu}\affiliation{\bama}
\author{R.~J.~Stefanski}\affiliation{\fnal}
\author{M.~Sung}\affiliation{\lsu}
\author{H.~A.~Tanaka}\altaffiliation{Present Address: \triumf}\affiliation{\princeton}
\author{R.~Tayloe}\affiliation{\indiana}
\author{M.~Tzanov}\affiliation{\colorado}
\author{R.~Van~de~Water}\affiliation{\lanl}
\author{M.~O.~Wascko}\altaffiliation{Present Address: \imperial}\affiliation{\lsu}
\author{D.~H.~White}\affiliation{\lanl}
\author{M.~J.~Wilking}\affiliation{\colorado}
\author{H.~J.~Yang}\affiliation{\umich}
\author{G.~P.~Zeller}\affiliation{\columbia}\affiliation{\lanl}
\author{E.~D.~Zimmerman}\affiliation{\colorado} 

\collaboration{MiniBooNE Collaboration}\noaffiliation

\date{\today}

\begin{abstract}
Using a cleanly tagged data sample of $\nu_\mu$ charged current events, it is demonstrated that the rate at which such events are mis-identified as $\nu_e$'s is accurately simulated in the MiniBooNE $\nu_\mu \rightarrow \nu_e$ analysis. Such mis-identification, which could arise from muon internal bremsstrahlung, is decisively ruled out as a source of the low energy electron-like events reported in the MiniBooNE search for $\nu_\mu \rightarrow \nu_e$ oscillations.   This refutes the conclusions of a recent paper which postulates that hard bremsstrahlung could form a substantial background to the MiniBooNE $\nu_e$ sample.
\end{abstract}

\maketitle


The MiniBooNE Collaboration has reported the results of a search for
$\nu_\mu \rightarrow \nu_e$ oscillations at $\Delta m^2 \sim 1$ eV$^2$.
In this search, no significant excess of events was observed above background for
reconstructed neutrino energies above 475 MeV, but $96 \pm 17 (stat)
\pm 20 (sys)$ excess events were reported between 300 and 475 MeV 
\cite{oscpaper}. The data are not consistent with two-neutrino
oscillations and the source of the excess is, at present, undetermined.

\begin{figure}[b]
\includegraphics[height=1.1in]{{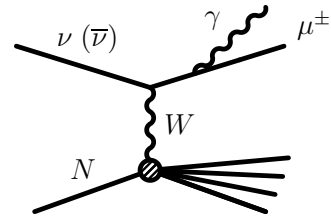}}
\caption{Feynman diagram for muon internal bremsstrahlung.}
\label{fig:diagram}
\end{figure}

A recent paper \cite{Arie} suggests that the source of the low energy excess
is muon internal bremsstrahlung associated with the $\nu_\mu$ charged
current quasi-elastic (CCQE) interaction, $\numu \, n \rightarrow \mu^- \, p +
\gamma$, as depicted in
Fig~\ref{fig:diagram}.  The issue of bremsstrahlung of muon neutrino CCQE events was 
raised in an earlier paper \cite{Efrosinin} and brought to MiniBooNE's attention in the course
of the oscillation analysis \cite{beacom}, prompting the study presented here.

Reference \cite{Arie} asserts that $\numu \, n \rightarrow \mu^- \, p +
\gamma$ events can be mis-classified as $\nue$ CCQE 
signal events if the final state muon is below \kov threshold, since the radiated
photon may imitate a final state electron. Even when the muon escapes 
direct detection, however, its presence may be revealed by the Michel electron from the 
muon decay. Making use of this Michel electron tag, the present paper demonstrates that 
the misidentification of $\numu$ events 
caused by muon internal bremsstrahlung cannot
be the source of the observed low energy excess and is not a significant
background to the MiniBooNE $\nu_\mu \rightarrow \nu_e$ oscillation search.
Rather than relying on Monte Carlo simulation of this
process, data are used to directly constrain the contribution of
the muon bremsstrahlung diagram. The study
was conducted prior to the unblinding of the data for the oscillation analysis 
and the results were incorporated into the estimated backgrounds at that time. Because the study showed that the background to the oscillation analysis from muon internal bremsstrahlung was extremely small, it was deemed unnecessary to add such internal radiative effects to the simulation.  The analysis makes use of an event
sample in which the presence of a muon is tagged strictly by the
presence of the Michel electron from the decay of the muon:

\begin{displaymath}
  \nu_\mu+n \rightarrow \mu^{-}+p, \hspace{0.2in}
  \mu^{-} \rightarrow e^{-} + \nu_\mu + \bar{\nu_e}.
\end{displaymath}


\begin{figure}
\includegraphics[width=\columnwidth]{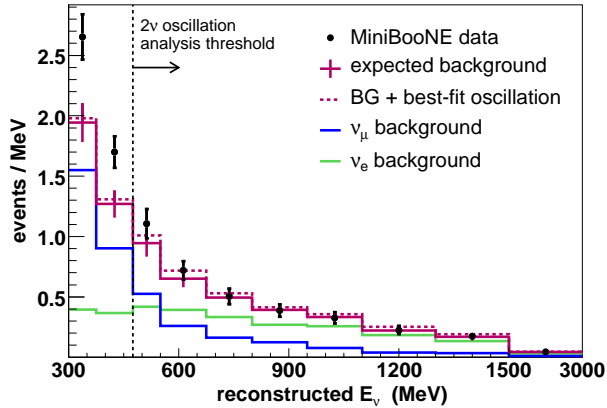}
\caption{Reconstructed neutrino energy distribution of {\em one} subevent events 
passing $\nue$ selection cuts.  The data are shown as points with statistical errors and
the Monte Carlo prediction is shown as a histogram with systematic errors. 
This figure is a reproduction of Fig.~2 in \cite{oscpaper}. }
\label{fig:excess}
\end{figure}

\begin{figure}
\includegraphics[width=\columnwidth]{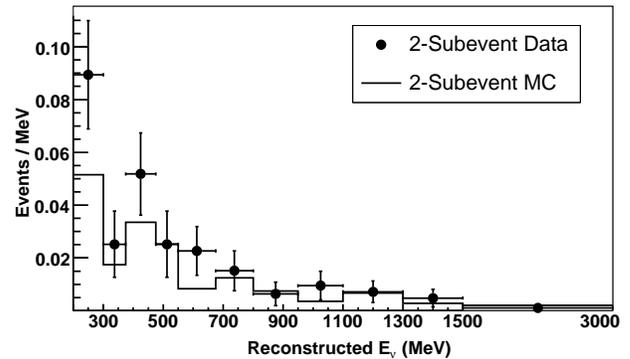}
\caption{Reconstructed neutrino energy distribution of {\em two} subevent events 
passing $\nue$ selection cuts. The data are shown as points with statistical errors and
the Monte Carlo prediction is shown as a histogram. No systematic uncertainty has been evaluated for this sample. Normalization of both data and 
Monte Carlo is to the rate of $\nu_\mu$ CCQE events with no observed Michel.  If muon internal bremsstrahlung were the source of the low energy excess in the oscillation sample (Fig.~\ref{fig:excess}), then the same excess would also be observed in this figure, but is not.}
\label{fig:TwoSEPassingCutsRenormed}
\end{figure}

The MiniBooNE detector and trigger \cite{oscpaper} are particularly well suited 
for such identification.
The MiniBooNE trigger creates a 19.2 $\mu$s time window surrounding the 1.6
$\mu$s beam spill. When events are reconstructed, periods of time
in which light is produced in the tank are identified as ``subevents''.
Subevents are separated by looking for gaps between PMT hit times larger than 10 ns,
and are typically $\sim$100 ns in length.

In 82\% of the cases where a muon is contained in the detector, a
second subevent from the Michel electron is produced.  The 18\%
without a second subevent splits into 8\% that result from $\mu^-$ 
capture in oil (this rate has been separately measured \cite{muon_capture}), 
2\% where simulation predicts the Michel electron creates too few PMT hits 
to be clearly seen ($<$10 hit PMTs), and 8\% where the muon decays sufficiently 
quickly that the decay cannot be time-resolved from the initial interaction.

To study the rate at which $\numu$ charged current interactions are mis-identified as $\nue$'s,
 events with two subevents are first selected. The two subevents must be separated by 
at least 1500 ns to avoid any instrumental effects causing the second subevent. 
No cuts are placed on the first subevent, but the second subevent must have fewer than 6
 veto hits to ensure containment and reject cosmic muons and
fewer than 200 main tank hits to ensure an energy consistent 
with a Michel electron. Having created, in this way, a sample of events tagged as being 
from a $\numu$ charged current interaction, the second subevent and its hits are discarded. 
The full set of $\nue$ selection cuts, identical to those used in the oscillation result, 
are then applied to this artificial one subevent sample. The resulting sample is
 a direct, {\em in situ} measurement of the bulk of the $\numu$ charged current
 contribution to the background in the $\nue$ sample, including that 
from muon internal bremsstrahlung. The only $\numu$ charged current background
 component this measurement misses is that due to muon decay so rapid that the Michel electron cannot be time separated from the parent muon. This background is constrained and checked in other ways.

The $\nue$ selection cuts \cite{oscpaper} include precuts to isolate a clean neutrino 
event sample: the event must have just one subevent within the 1.6 $\mu$s beam window 
and have fewer than 6 veto hits to remove incoming cosmic rays and exiting muons from 
neutrino events in the detector. Each event must also have more than 200 tank hits to
reject Michel electrons from stopped cosmic rays which can enter the
tank prior to the trigger window and therefore avoid the veto.

After the precuts, the same ``track-based" algorithm used in the oscillation analysis 
\cite{oscpaper} reconstructs the vertex position, angle, energy, and time of the event,
assuming the light comes from an extended, straight line source.
The vertex and projected track endpoint must lie within the fiducial volume of
the detector and the visible energy in the tank must be $E_{vis}>140$ MeV.

Events are identified as $\nu_e$ CCQE based on
likelihoods which are built from phototube charge and time
probability distribution functions. For each event, using a
single-track hypothesis, the likelihood is calculated that it is an electron ($L_e$) or a muon
($L_\mu$). The event is then reconstructed under a two track
hypothesis, where the invariant mass is forced to be 135 MeV, and a
$\pizero$ likelihood ($L_\pi$) is formed. Finally, the event is
reconstructed by finding the best two track fit allowing the invariant mass
to float. This yields a best-fit mass ($M_{\gamma\gamma}$).  
Visible energy-dependent cuts on $\log(L_e/L_\mu)$, $\log(L_e/L_\pi)$ 
and $M_{\gamma\gamma}$ are then applied to isolate a $\nue$ CCQE signal sample. 
The neutrino energy is then determined using the reconstructed 
lepton energy and direction and assuming the interaction is $\nue$ CCQE.

Fig.~\ref{fig:excess} shows the spectrum of the originally-published $\nue$ 
candidate sample subject to this selection \cite{oscpaper}.
The excess of events between 300 and 475 MeV is clearly visible. 
Fig.~\ref{fig:TwoSEPassingCutsRenormed} shows the results of the same cuts 
applied to the tagged $\numu$ data sample. In this case, the
normalization has been adjusted to correct for the Monte Carlo determined rates 
of one subevent and two subevent $\numu$ charged current interactions and for the
requirement that the two subevents be separated by more than 1500 ns.
Fig.~\ref{fig:TwoSEPassingCutsRenormed} is therefore a direct prediction and 
measurement of the $\numu$ charged current contribution to Fig.~\ref{fig:excess}. 
The Monte Carlo prediction for the rate at which these processes 
pass the $\nue$ selection agrees well with the data. Furthermore, the 
abscissa of these two figures shows that $\numu$ charged current processes account 
for only a tiny fraction of the background in Fig.~\ref{fig:excess}. Muon internal 
bremsstrahlung (or any $\numu$ charged 
current process that can lead to muon decay at rest) therefore cannot be the source of the 
low energy data excess.


Muon bremsstrahlung
events should be efficiently rejected by the $\nue$ selection. Muon tracks are considerably longer than
electron tracks for the same visible energy in the MiniBooNE detector.
For events where the muon energy lies above \kov threshold, the additional track length 
leads to a significantly different charge and hit
structure, particularly in the center of the \kov ring. One would
therefore expect the presence of the muon to pull the charge and time
likelihoods of the events away from an ``electron-like" hypothesis.


In this paper, $\numu$ charged current events are identified through
the presence of a Michel electron and then subjected to the MiniBooNE 
oscillation analysis cuts.  The rate of misidentification of these events as $\nue$'s is
accurately modeled by the Monte Carlo and is not the source of the
low energy excess in MiniBooNE.

It should be noted that this study says nothing about the total rate of muon 
internal bremsstrahlung in MiniBooNE, just the rate at which this process 
occurs in the $\nue$ sample. It may be possible to relax or adjust the 
$\nue$ selection cuts to make a measurement of muon internal bremsstrahlung
at some time in the future.

\bigskip

\begin{acknowledgments}
We acknowledge the support of Fermilab, the Department of Energy,
and the National Science Foundation in the construction, operation, and 
data analysis of the MiniBooNE experiment. 
\end{acknowledgments}


\end{document}